\documentclass[12pt]{article}
\usepackage{epsfig}
\usepackage{xcolor}

\newcommand{\bea}{\begin{eqnarray}}
\newcommand{\eea}{\end{eqnarray}}
\newcommand{\be}{\begin{equation}}
\newcommand{\ee}{\end{equation}}

\newcommand{\as}{\alpha_s}
\newcommand{\asMZ}{\alpha_s(M^2_Z)}
\newcommand{\ar}{a_s}

\begin{document}

\begin{center}
  {\bf $\alpha_s$  from DIS data
   with  large $x$ resummations
  }

\vskip 0.5cm

A.V. Kotikov$^{a,b}$,
B.G.~Shaikhatdenov$^b$, N.S. Korchagin$^a$, and P. Zhang$^a$


\vskip 0.5cm
${}^a$ School of Physics and Astronomy, Sun Yat-sen University, Zhuhai 519082, China\\
${}^b$ Joint Institute for Nuclear Research, Russia
\end{center}


\begin{abstract}
  The deep inelastic scattering data on the nucleon $F_2$ structure function, accumulated by BCDMS, SLAC and NMC collaborations
  in fixed-target experiments, are analyzed in the non-singlet approximation within the frameworks of both conventional
   $\overline{\mathbf{MS}}$ scheme as well as those with resummations of logarithms at large Bjorken $x$ values.
  The use of the latter is important because they greatly modify the values of the twist four corrections
   while leaving a strong coupling constant almost intact.
\end{abstract}

$PACS:~~12.38~Aw,\,Bx,\,Qk$\\

{\it Keywords:} Deep inelastic scattering; structure functions;
QCD coupling constant; NNLO level; twist-four corrections.

\section{ Introduction }

Currently, the accuracy of the data for the structure functions (SFs) of deep inelastic scattering (DIS)  allows us to study
simultaneously the $Q^2$ dependence of logarithmic corrections based on QCD and power (nonperturbative) corrections
(see, for example,~\cite{Beneke} and references therein).

To date an achieved accuracy in most perturbative calculus on the market is at the level of the next-to-next-to-leading order (NNLO)
(see, for example,~\cite{PKK}-\cite{Kotikov:2015zda} and references cited therein).
However, relevant articles have recently been published in which the QCD analysis of DIS SFs
was carried out up to the next-to-next-to-next-to-leading order (NNNLO) of perturbative approach~\cite{Blumlein:2008kz}.

This paper is closely related to~\cite{Shaikhatdenov:2009xd,Kotikov:2015zda} which were devoted to similar studies, with the main
difference in that here we work within the framework of schemes whose application leads to effective resummation
of logarithms at large $x$ that contribute to the Wilson coefficient functions.
This is the so-called DIS
\footnote{The application of DIS scheme was also done in the short paper \cite{Kotikov:2022vlo}.}
~\cite{DIS} scheme, $W^2\sim Q^2(1-x)$-like
\footnote{The $W^2$-evolution is meant to incorporate an effect of soft emission \cite{Sterman:1986aj,Catani:1989ne}.}
evolution, and the Grunberg effective charge method~\cite{Grunberg}.
Thus we analyze the experimental data on the $F_2(x,Q^2)$ structure function collected by SLAC, NMC and BCDMS
collaborations~\cite{SLAC1}--\cite{BCDMS1} at the NNLO level of massless perturbative QCD.

Just like in our previous papers~\cite{Shaikhatdenov:2009xd,Kotikov:2015zda,KK2001}, the SF $F_2(x,Q^2)$ is
represented as a sum of the leading twist $F_2^{\rm pQCD}(x,Q^2)$ and twist four terms denoted by $\tilde h_4(x)$
\footnote{Further on, the superscripts {\rm pQCD,~LT} denote the twist two approximation with and without 
target mass corrections (see, for example, \cite{KK2001}).}:
\be
F_2(x,Q^2)=F_2^{\rm pQCD}(x,Q^2)\left(1+\frac{\tilde h_4(x)}{Q^2}\right)\,.
\label{1.1}
\ee

\section{ Theoretical intro to the analysis }

Here we briefly describe some aspects of theoretical part of our analyses.
More elaborate description can be found in~\cite{KK2001,Shaikhatdenov:2009xd}.
Note that in the region of large $x$ values the gluons do nearly not contribute and
$Q^2$-evolution of the twist-two DIS SF $F_2(x,Q^2)$ is almost completely determined by its so-called
non-singlet (NS) part.

In this approximation, there is a direct relationship between the (Mellin) moments of the DIS SF $F_2(x,Q^2)$
and the corresponding moments of the NS parton distribution function (PDF) $\tilde{f}(x,Q^2)$
\footnote{Unlike the standard case, here PDF is multiplied by $x$.}
\be
M(n,Q^2) ~=~\int_0^1 dx x^{n-2} F^{\rm LT}_{2}(x,Q^2),~~
{\bf f}(n,Q^2) ~=~\int_0^1 dx x^{n-2}\, \tilde{f}(x,Q^2) 
\label{Moments}
\ee
as follows
\be
M(n,Q^2) = R_{\rm NS}(f)\cdot C(n,\ar(Q^2))\cdot {\bf f}(n,Q^2)\,,
\label{3.a}
\ee
where the strong coupling constant
\be
\ar(Q^2)=\frac{\alpha_s(Q^2)}{4\pi} \label{as}
\ee
and $C(n,\ar(Q^2))$  denotes a Wilson coefficient function.
The constant $R_{NS}(f)$ depends on weak and electromagnetic charges and is fixed as $R_{NS}(f=4)=1/6$
\cite{Buras}. Here and below $f$ is the number of active quark flavors.

\subsection{Strong coupling constant}

The strong coupling constant is determined from the corresponding equation of the renormalization group.
At the NLO level, it is obtained as follows
\bea \label{1.coA}
\frac{1}{a_{1}(Q^2)} - \frac{1}{a_{1}(M_Z^2)} +
b_1 \ln{\left[\frac{a_{1}(Q^2)}{a_{1}(M_Z^2)}
\frac{(1 + b_1a_{1}(M_Z^2))}
{(1 + b_1a_{1}(Q^2))}\right]}
= \beta_0 \ln{\left(\frac{Q^2}{M_Z^2}\right)}\,,
\eea
where hereafter
\be
a_{1}(Q^2)=\ar^{\rm NLO}(Q^2),~~a_{2}(Q^2)=\ar^{\rm NNLO}(Q^2) \, .
\label{ai}
\ee
At the NNLO level, the coupling constant is determined from the following equation
\bea \label{1.co}
\frac{1}{a_2(Q^2)} - \frac{1}{a_2(M_Z^2)} &+&
b_1 \ln{\left[\frac{a_2(Q^2)}{a_2(M_Z^2)}
\sqrt{\frac{1 + b_1a_2(M_Z^2) + b_2a_2^2(M_Z^2)}
{1 + b_1a_2(Q^2) + b_2a_2(Q^2)}}\right]} \\ \nonumber
&+& \left(b_2-\frac{b_1^2}{2}\right)\cdot
\Bigl(I(a_s(Q^2))- I(a_s(M_Z^2))\Bigr) = \beta_0 \ln{\left(\frac{Q^2}{M_Z^2}\right)}\,.
\eea
The expression for $I$ looks like
$$
I(a_s(Q^2))=\cases{
\displaystyle{\frac{2}{\sqrt{\Delta}}} \arctan{\displaystyle{\frac{b_1+2b_2a_2(Q^2)}{\sqrt{\Delta}}}} &for $f=3,4,5; \Delta>0$,\cr
\displaystyle{\frac{1}{\sqrt{-\Delta}}}\ln{\left[
\frac{b_1+2b_2a_2(Q^2)-\sqrt{-\Delta}}{b_1+2b_2a_2(Q^2)+\sqrt{-\Delta}}
\right]}&for $f=6;\quad\Delta<0$, \cr
}
$$
where $\Delta=4b_2 - b_1^2$ and $b_i=\beta_i/\beta_0$ are taken from the QCD $\beta$-function:
$$
\beta(\ar) ~=~ -\beta_0 \ar^2 - \beta_1 \ar^3 - \beta_2 \ar^4 +\ldots \,.
$$

\subsection{$Q^2$-dependence of SF moments}

The Wilson coefficient function $C(n,\ar(Q^2))$ is expressed in terms of coefficients $B_j(n)$ (hereafter (j=1,2))
(see, for example, \cite{Shaikhatdenov:2009xd})
\footnote{For odd $n$ values, $B_j(n)$ and $Z_j(n)$ coefficients can be obtained by using the analytical continuation~\cite{KaKo}.}
\be
C(n, \ar(Q^2)) = 1 
+ \ar(Q^2) B_{1}(n)
+ \ar^2(Q^2) B_{2}(n) + {\cal O}(\ar^3)\,.
\label{1.cf}
\ee

The $Q^2$-evolution of PDF moments can be found within the framework
of perturbative QCD (see, for example, \cite{Buras}):
\be
\frac{{\bf f}(n,Q^2)}{{\bf f}(n,Q_0^2)}=\left[\frac{\ar(Q^2)}{\ar(Q^2_0)}\right]^{d(n)}
\frac{h(n, Q^2)}{h(n, Q^2_0)}
\,, 
\label{3}
\ee
where
\be\label{hnns}
d(n)=\frac{\gamma_{0}(n)}{2\beta_0},~~
h(n, Q^2)  = 1 + \ar(Q^2) Z_{1}(n) + \ar^2(Q^2) Z_{2}(n)
+ {\cal O}\left(\ar^3\right)\,,
\ee
and
\begin{eqnarray}
Z_{1}(n) &=& \frac{1}{2\beta_0} \biggl[ \gamma_{1}(n) -
\gamma_{0}(n)\, b_1\biggr]\,, \nonumber \\
Z_{2}(n)&=& \frac{1}{4\beta_0}\left[
\gamma_{2}(n)-\gamma_{1}(n)b_1 + \gamma_{0}(n)(b^2_1-b_2) \right]
+  \frac{1}{2} Z^2_{1}(n)
\,
\label{3.21}
\end{eqnarray}
are combinations of the NLO and NNLO anomalous dimensions $\gamma_{1}(n)$ and $\gamma_{2}(n)$.

For large $n$ (this corresponds to large values of $x$), the coefficients $Z_{j}(n)\sim\ln n$ and $B_{j}(n)\sim\ln^{2j}n$.
Thus, the coefficients $B_{j}(n)$ can lead to potentially large contributions and, therefore, they should be resummed.
This will be done in Section 3, mainly with the appropriate choice of factorization scale.

\subsection{Factorization $\mu_F$ and renormalization $\mu_R$ scales}

Here we intend to consider dependence of the above results on the factorization $\mu_F$ and
renormalization $\mu_R$ scales caused (see, for example, ~\cite{Shaikhatdenov:2009xd})
by truncation of a perturbative series. The modification is achieved
by replacing $\ar(Q^2)$ in Eqs.~(\ref{3.a}) and (\ref{3}) by the expressions in which the scales were taken
into account as follows: $\mu^2_F= k_F Q^2,~~\mu^2_R=k_R \mu^2_F = k_R k_F Q^2$.

Then, Eq.~(\ref{3.a}) takes the form:
\be
M(n,Q^2) = R_{NS}(f) \cdot \hat{C}(n, \ar(k_F Q^2))
\cdot {\bf f}(n,k_F Q^2), \nonumber
\ee
and Eq.~(\ref{3}) gets replaced by
\be
\frac{{\bf f}(n, k_F Q^2)}{{\bf f}(n,k_F Q_0^2)}
=\bigg[\frac{\ar(k_F k_R Q^2)}
{\ar(k_F k_R Q^2_0)}\bigg]^{d(n)}
\cdot \frac{\hat{h}(n, k_Fk_RQ^2)}{\hat{h}(n,  k_Fk_RQ^2_0)}
\label{3l}
\ee

The functions $\hat{C},\hat{h}$  are to be obtained from $C$, $h$ by modifying r.h.s. of Eqs.~(\ref{1.cf}) and
(\ref{hnns}) as follows:\\
in Eq.~(\ref{1.cf})
\bea
\ar(Q^2) &\to& \ar(k_F Q^2)\,,
B_{1}(n) \to B_{1}(n) + \frac{1}{2}\gamma_{0}(n) \ln{k_F}\,, \label{bnlo} \\
B_{2}(n) &\to& B_2(n) + \frac{1}{2}\gamma_{1}(n) \ln{k_F}
+\left(\frac{1}{2}\gamma_{0} + \beta_0\right)B_{1}\ln{k_F} +
\frac{1}{8}\gamma_{0}\left(\gamma_{0} + 2\beta_0\right)\ln^2{k_F}\,,
\label{coeffun}
\eea
and in Eq.~(\ref{hnns})
\bea
\ar(Q^2) &\to& \ar(k_F k_R Q^2),~~ \ar(Q_0^2) \, \to \, \ar(k_F k_R Q_0^2),
\nonumber \\ \nonumber
Z_{1}(n) &\to& Z_{1}(n) + \frac{1}{2}\gamma_{0}(n)\ln{k_R}, \\
Z_{2}(n) &\to& Z_{2}(n) + \frac{1}{2}\gamma_{1}(n)\ln{k_R}
+ \frac{1}{2}\gamma_{0}(n)Z_{1}\ln{k_R}+
\frac{1}{8}\gamma_{0}\left(\gamma_{0} + 2\beta_0\right)\ln^2{k_R}\,.\nonumber
\eea

\section{Schemes with large $x$ resummations}

For large values of $x$ (and $n$), i.e. for $x\to 1$ (i.e. for $n\to\infty$ for Mellin moments), the coefficients $B_j(n)$
of the coefficient function $C(n)$ have the asymptotics $\ln^{2j}(n)$ and thus the most important terms should be
summed up. One of the most popular resummation procedures is the Catani--Trentadue one~\cite{Catani:1989ne},
in which the sum of the most important terms translates them to the exponent's argument.

Here we consider an alternative possibility in which the resumming most important terms changes the strong coupling argument,
which then becomes $n$-dependent.
In this section, we will look at three different schemes. Two of them are the DIS-scheme \cite{DIS} and $W^2$-evolution, which contain resummations in $n$- and $x$-spaces, respectively.
Here $W^2=(p+q)^2$ is an effective mass of the DIS process. In the massless limit, it can be represented as $W^2=Q^2(1-x)/x$,
and therefore its usage reduces the largest powers of $\ln(1-x)$ of the coefficient function $C(n)$. These schemes are set up by
changing the factorization scale. The third scheme is the well-known Grunberg effective charge method, which is characterized
by modifying the factorization and renormalization scales and,  beyond NLO, by a change in the coefficients
$\beta_j$ $(j\geq2)$ of the QCD $\beta$ function.

Now we will look at these three schemes separately.

\subsection{DIS scheme}

Let us consider the case of the so-called DIS-scheme~\cite{DIS} (it was alternatively called the $\Lambda_n$-scheme \cite{Lambda_n}),
where NLO corrections to the Wilson coefficients are completely compensated by changing the factorization scale.

\subsubsection{NLO}

In this order, the strong coupling constant changes as follows
\be
a_s(Q^2) \to a_s(k_{\rm DIS}(n)Q^2)\equiv a_{1,n}^{\rm DIS}(Q^2),~~
k_{\rm DIS }(n)=\exp\left(\frac{-2B_{1}(n)}{\gamma_0(n)}\right) =\exp\left(\frac{-r^{\rm DIS }_{1}(n)}{\beta_0}\right) \, ,
\label{kDIS.NLO}
\ee
where
\be
r^{\rm DIS }_{1}(n)=\frac{2B_{1}(n)\beta_0}{\gamma_0(n)}
~~~~\mbox{and}~~~
 B_1(n) \to B^{\rm DIS}_{1} =0,
\label{oBDI.NLO}
\ee
i.e. $\hat{C}(n, a_{1,n}^{\rm DIS}(Q^2))=1+ {\cal O}((a_{1,n}^{\rm DIS})^2)$.

The NLO coupling constant $a^{\rm DIS}_{1,n}(Q^2)$ obeys the following equation
\bea \label{1.coA.DIS}
&&\frac{1}{a^{\rm DIS}_{1,n}(Q^2)} - \frac{1}{a_{1}(M_Z^2)} +
b_1 \ln{\left[\frac{a^{\rm DIS}_{1,n}(Q^2)}{a_{1}(M_Z^2)}
\frac{(1 + b_1a_{1}(M_Z^2))}
{(1 + b_1a^{\rm DIS}_{1,n}(Q^2))}\right]}= \beta_0 \ln{\left(\frac{k_{\rm DIS}(n)Q^2}{M_Z^2}\right)}\nonumber \\
&&\hspace{1cm}  = \beta_0 \ln{\left(\frac{Q^2}{M_Z^2}\right)}-r^{\rm DIS}_{1}(n)\,,
\eea
which can be obtained from Eq. (\ref{1.coA}) by substituting $ Q^2 \to k_{\rm DIS}(n)Q^2$.

Hereafter the condition that coupling constants in all schemes coincide at $Q^2=M_Z^2$ is satisfied.

\subsubsection{NNLO}

At this level of accuracy, we have to use Eqs. (\ref{kDIS.NLO}) and (\ref{oBDI.NLO}) and, in addition, the NNLO Wilson coefficient $B_2(n)$ that is modified as follows
\be
B_2(n) \to B^{\rm DIS}_2(n)= B_2(n)-\left(\frac{1}{2}+\frac{\beta_0}{\gamma_0(n)}\right)\, B^2_1(n)-
\frac{\gamma_1(n)}{\gamma_0(n)}\, B_1(n)\,.
\label{oBDI.NNLO}
\ee
This leads to the complete cancellation of the larger terms $\sim\ln^4(n)$ and $\sim\ln^3(n)$ in $B^{\rm DIS}_2(n)$.

We have
\be
\hat{C}(n, a_{2,n}^{\rm DIS}(Q^2))=1+  B^{\rm DIS}_2(n)\left(a_{2,n}^{\rm DIS}\right)^2+{\cal O}((a_{2,n}^{\rm DIS})^3)
\label{C2DIS}
\ee
and the NNLO coupling constant $a^{\rm DIS}_{2,n}(Q^2)$ obeys the equation
\bea \label{1.coB.DIS}
\frac{1}{a^{\rm DIS}_{2,n}(Q^2)} - \frac{1}{a_2(M_Z^2)} &+&
b_1 \ln{\left[\frac{a^{\rm DIS}_{2,n}(Q^2)}{a_2(M_Z^2)}
\sqrt{\frac{1 + b_1a_2(M_Z^2) + b_2a_2^2(M_Z^2)}
{1 + b_1a^{\rm DIS}_{2,n}(Q^2) + b_2(a^{\rm DIS}_{2,n})^2(Q^2)}}\right]} \\ \nonumber
&& \hspace{-2cm} +\left(b_2-\frac{b_1^2}{2}\right)\times
\Bigl(I(a^{\rm DIS}_{2,n}(Q^2))-I(a_s(M_Z^2))\Bigr) = \beta_0 \ln{\left(\frac{Q^2}{M_Z^2}\right)}-r^{\rm DIS}_{1}(n)\,,
\eea
which can be obtained from Eq. (\ref{1.co}) by substituting $ Q^2 \to k_{\rm DIS}(n)Q^2$.

 \vskip 0.5cm

 Thus, in DIS scheme, the NLO coefficient $B_1(n)$ is exactly compensated by changing the strong coupling argument.
 Moreover, for large values of $n$, the NNLO coefficient $B^{\rm DIS}_2(n)$ contains only terms $\sim \ln^2n$ whilst the
 leading terms of the form $\sim \ln^4n$ and $ \sim \ln^3n$ are cancelled out.

 \subsection{$W^2$
   scheme}

 Here we use the fact that in NLO, the basic contibution in the $x$-space looks like
 \footnote{This property is also important for $N=4$ Super Yang--Mills \cite{Bianchi:2013sta}.}~\cite{Bardeen:1978yd,Kubar-Andre:1978eri}
\bea
\tilde{B}^{W}_{1}(x) = - \frac{1}{2}\tilde{\gamma}_{0}(x) \ln{\overline{k}_W(x)}\,,~~\overline{k}_W(x)=\frac{1-x}{x}\,,
\label{basic}
\eea
where the quantities marked by tilde do not contain the contribution coming from sum rules.
Our standard Wilson coefficient $B_{1}(n)$
and $\gamma_{0}(n)$
are obtained from
$\tilde{B}_{1}(n)$ and
$\tilde{\gamma}_{0}(n)$ by adding the sum rule conditions as follows
\be
B_{1}(n)=\tilde{B}_{1}(n)-\tilde{B}_{1}(n=1),~~\gamma_{0}(n)=\tilde{\gamma}_{0}(n)-\tilde{\gamma}_{0}(n=1)\,,
\label{BtB}
\ee
with 
\be
\tilde{B}_{1}(n)=\int_0^1 \, dx x^{n-1}\, \tilde{B}_{1}(x),~~ \tilde{\gamma}_{0}(n)=\int_0^1 \, dx x^{n-1}\,
\tilde{\gamma}_{0}(x)\,.
\label{MellinB}
\ee

The scale $\overline{k}_W(x)$ is related to the massless limit of the effective mass $W$ of a photon--proton cluster
\be
W^2=Q^2\frac{1-x}{x}+M_P^2\,,
\label{W^2}
\ee
where $M_p$ is a proton mass. In the massless limit ($M_P=0$) $W^2=\overline{k}_W(x)\,Q^2$; thus, in this subsection,
we use the scale $\overline{k}_W(x)$ in the calculations.

\subsubsection{NLO}

In this order,
\be
a_s(Q^2) \to a_s(k_{W}(n)Q^2)\equiv a_{1,n}^{W}(Q^2),~~
k_{W}(n)=exp\left(\frac{-2B^{W}_{1}(n)}{\gamma_0(n)}\right) =exp\left(\frac{-r^{W}_{1}(n)}{\beta_0}\right) \, ,
\label{kW.NLO}
\ee
where
\be
r^{W}_{1}(n)=\frac{2B^{W}_{1}(n)\beta_0}{\gamma_0(n)}=\frac{B^{W}_{1}(n)}{d(n)}
\label{r1W}
\ee
and
\be
\hat{C}(n, \ar(Q^2)) = 1 +  a_{1,n}^{W}(Q^2)
\hat{B}^{W}_{1}(n)
+ {\cal O}((a_{1,n}^{W})^2)\,.
\label{1DI.cf_W2}
\ee
Here
\be
\hat{B}^{W}_1(n) = B_1(n)- B^{W}_{1}(n) ,
\label{oBW.NLO}
\ee
with
\be
B^{W}_{1}(n) = \tilde{B}^{W}_{1}(n)-\tilde{B}^{W}_{1}(n=1),~~
\tilde{B}^{W}_{1}(n)=\int_0^1 \, dx x^{n-1}\, \tilde{B}^{W}_{1}(x),~~
 \label{oBW.NLO}
\ee
and $\tilde{B}^{W}_{1}(x)$
is defined above in Eq.~(\ref{basic}).

Then, we have ($C_F=(N^2-1)/(2N)$ for SU(N) group)
\bea
&&B^{W}_{1}(n) =  2C_F \left[S_1^2(n) + S_2(n) -
  \frac{1}{n(n+1)}
    \,S_1(n) -\frac{7}{4}
    \right] \, ,
\label{oBWa.NLO} \\
&&\hat{B}^{W}_{1}(n) = 2C_F \left[-2 S_2(n) +\frac{3}{2}
  \,S_1(n) - \frac{11}{4} + \frac{3}{2n} + \frac{2}{n+1}
  + \frac{1}{n^2}  - \frac{1}{(n+1)^2} \right]  .
\label{hBWa.NLO}
\eea

At large $x$ values (and at large $n$ values, respectively), when ($\zeta_i$ are Euler $\zeta$-functions)
\be
S_1(n)=
\ln n + \gamma_{\rm E} + O(1/n),~~ S_i(n)=\zeta_i+ O(1/n),~~ (i=2,...),
\label{Si_Large}
\ee
we have
\be
\hat{B}^{W}_{1}(n) = 2C_F \left[-2 \zeta_2 + \frac{3}{2}\,\ln n - \frac{11}{4} 
  + O(n^{-1}) \right] \sim 3C_F\,
\ln n +  O(n^{0})\,
\label{hB1_Large}
\ee
and the results
look very similar to those obtained in the previous subsection.

Note that the NLO coupling constant $a^{W}_{1,n}(Q^2)$ obeys
Eq. (\ref{1.coA.DIS})
where the replacement $r^{\rm DIS}_{1}(n) \to r^{W}_{1}(n)$ is done.

\subsubsection{NNLO}

Here,
\be
a_s(Q^2) \to  a_s(k_{W}(n)Q^2)\equiv a^{W}_{2,n}(Q^2),~~
\label{kW.NNLO}
\ee
and
\be
\hat{C}(n, \ar(Q^2)) = 1 + a^{W}_{2,n}(Q^2)
\hat{B}^{W}_{1}(n) + {\left(a^{W}_{2,n}(Q^2)\right)}^2
\hat{B}^{W}_{2}(n)
+ {\cal O}((a_{2,n}^{W})^3)\,,
\label{1DI.cf_W2}
\ee
with $B^{W}_{1}(n) $ and $\hat{B}^{W}_{1}(n)$ given in Eqs.~(\ref{oBWa.NLO}) and (\ref{hBWa.NLO}), respectively, and
\be
\hat{B}^{W}_2(n)= B_2(n) -\left(\frac{1}{2}+\frac{\beta_0}{\gamma_0(n)}\right)\, B^{W}_1(n) \Bigl[2B_1(n)-B^{W}_1(n)\Bigr]
-\frac{\gamma_1(n)}{\gamma_0(n)}
\, B^{W}_1(n)\, .
\label{oBW.NLO}
\ee

Since the DIS-scheme and $W^2$-evolution are close to each other,
it is convenient to express $\hat{B}^{W}_2(n)$ through ${B}^{\rm DIS}_2(n)$:
\be
\hat{B}^{W}_2(n)= B^{\rm DIS}_2(n)+\left(\frac{1}{2}+\frac{\beta_0}{\gamma_0(n)}\right)\,  {\left(\hat{B}^{W}_1(n)\right)}^2
+\frac{\gamma_1(n)}{\gamma_0(n)}
\, \hat{B}^{W}_1(n)\, ,
\label{oBW.NLO.1}
\ee
where $ B^{\rm DIS}_2(n)$ is given in Eq.~(\ref{oBDI.NNLO}).

Note that the leading terms $\sim \ln^4(n)$ are cancelled out in $B^{W}_2(n)$.

The NNLO coupling $a^{W}_{2,n}(Q^2)$ obeys Eq. (\ref{1.coA.DIS})
where the replacement $r^{\rm DIS}_{1}(n) \to r^{W}_{1}(n)$ is carried out.

\vskip 0.5cm
Thus, in the $W^2$-evolution, for large  $n$ values
the NLO and NNLO coefficients $\hat{B}^{W}_1(n)$ and $\hat{B}^{W}_2(n)$  contain only the terms $\sim \ln n$ and
$\sim \ln^3n$, respectively. The most important terms are completely cancelled.\\

Note that $k_W(n)$ (or equally well $\overline{k}_W(x)$) can be multiplied by an additional factor $e^{\delta}$:
\be
\overline{k}_{\delta}(x)=e^{\delta}\,\overline{k}_W(x),~~k_{\delta}(n)=e^{\delta}\,k_W(n)\,.
\label{delta_k}
\ee

In the case when $\delta=-3/4$, we have
\be
\hat{B}^{(\delta=-3/4)}_{1}(n) = 2C_F \left[-2 S_2(n) - \frac{13}{8} + \frac{9}{4n} + \frac{5}{4(n+1)} + \frac{1}{n^2}
    - \frac{1}{(n+1)^2} \right]\,
\label{hBWa.NLO.d}
\ee
and $\hat{B}^{(\delta=-3/4)}_{1}(n) \sim O(n^0)$ at large $n$ values. The corresponding NNLO coefficient
$\hat{B}^{(\delta=-3/4)}_{2}(n) \sim \ln^2 n$ and it is seen that the case $\delta=-3/4$ is very close to the DIS-scheme.

Unfortunately, for several first values of $n$ starting with $n=2$, which are used in our fits (see Section 5),
$\hat{B}^{(\delta=-3/4)}_{1}( n)$ has large negative values, which worsens the quality of the fits. Therefore, we will not use
the case $\delta=-3/4$ in this study.

\section{Grunberg approach
}

In this subsection, we consider the Grunberg effective charge method, which is a fairly popular approach.
In a sense, it is closely related to the so-called scheme-invariant perturbation theory (SIPT),
which is as well widely in use (see, for example, Ref. \cite{PKK,Vovk}). In this approach, all contributions
beyond LO are completely canceled  by changes in the factorization and renormalization scales, while beyond
NLO it is also required to modify the coefficients $\beta_i$ ($i\geq 2$) of the QCD $\beta$ function.

In order to apply Grunberg approach, it is rather convenient to rewrite Eq.~(\ref{3.a}) as follows
\be
M_n(Q^2) =
\frac{\overline{C}(n,\ar(Q^2))}{\overline{C}(n,\ar(Q_0^2))} \cdot M_n(n,Q_0^2)\,,
\label{3.SI}
\ee
where
\be
\overline{C}(n,\ar(Q^2))=\ar^{d(n)}(Q^2)\, \Bigl[1+C_1 \ar(Q^2) + C_2 \ar^2(Q^2) + ...\Bigr]
\label{oC.SI}
\ee
contains contributions coming from both coefficient function $C(n,\ar(Q^2))$
and PDF evolution~(\ref{hnns}).
Indeed,
\be
C_{1}(n)=B_{1}(n)+Z_{1}(n),~~C_{2}(n)=B_{2}(n)+Z_{1}(n)B_{1}(n)+Z_{2}(n) \, ,
\label{Ci.SI}
\ee
where $Z_{i}(n)$ $(i=1,2)$ are shown in Eq. (\ref{3.21}).

The normalization $M_n(Q^2_0)$ is linked with ${\bf f}_{NS}(n,Q^2_0)$ as given in (\ref{3.a}) where $Q^2 \to Q^2_0$, i.e.
\be
M_n(Q^2_0) = R_{NS}(f)\cdot \overline{C}(n,\ar(Q^2_0))\cdot {\bf f}(n,Q^2_0)\,.
\label{3.a0}
\ee

\subsection{NLO}

In this order, the strong coupling constant is modified as follows:
\be
a_1(Q^2) \to a_1(k_{\rm SI}(n)Q^2)\equiv a_{1,n}(Q^2),~~
k_{\rm SI}(n)=exp\left(\frac{-2C_{1}(n)}{\gamma_0(n)}\right)=exp\left(\frac{-r_{1}(n)}{\beta_0}\right) \, ,
\label{kSI.NLO}
\ee
where
\be
r_{1}(n)=\frac{2C_{1}(n)\beta_0}{\gamma_0(n)}=\frac{C_{1}(n)}{d(n)}\,.
\label{r1}
\ee

With the above choice of the scale, we have
\be
C_1(n)= 0\,,~~
\mbox{i.e.}~~
\overline{C}(n)= a_{1,n}^{d(n)}(Q^2)\, \Bigl[1+O(a_{1,n}^2)\Bigr]\, .
\label{oCSI.NLO}
\ee

The NLO coupling $a_{1,n}(Q^2)$ obeys Eq.~(\ref{1.coA.DIS}),
where the replacement $r^{\rm DIS}_{1}(n) \to r_{1}(n)$ is done.

\subsection{NNLO}

Here,
\be
a_2(Q^2) \to a_2(k_{\rm SI}(n)Q^2))\equiv a_{2,n}(Q^2),~~
\label{kSI.NNLO}
\ee
where
the NNLO coupling $a_{2,n}(Q^2)$ obeys the following equation
\bea 
&&\frac{1}{a_{2,n}(Q^2)} - \frac{1}{a_2(M_Z^2)} +
b_1 \ln{\left[\frac{a_{2,n}(Q^2)}{a_2(M_Z^2)}
\sqrt{\frac{1 + b_1a_2(M_Z^2) + b_2a_2^2(M_Z^2)}
{1 + b_1a_{2,n}(Q^2) + \tilde{b}_2(n)a^2_{2,n}(Q^2)}}\right]}\nonumber \\ \nonumber
&&+ \left(\tilde{b}_2(n)-\frac{b_1^2}{2}\right)\cdot
\Bigl(\tilde{I}(a_{2,n}(Q^2))-\tilde{I}(0)\Bigr)-
\left(b_2-\frac{b_1^2}{2}\right)\cdot \Bigl(I(a_s(M_Z^2))-I(0)\Bigr)
 \\  &&= \beta_0 \ln{\left(\frac{Q^2}{M_Z^2}\right)}-r_{1}(n)\,,
 \label{1.coB.SI} \eea
 where
\be
 \tilde{I}(a_{2,n}(Q^2))=I(a_2(Q^2)\to a_{2,n}(Q^2), b_2 \to \tilde{b}_2(n))\,.
\label{tI}
 \ee
A modified factor is found to be
\be
\tilde{b}_2(n)=\frac{\tilde{\beta}_2(n)}{\beta_0},~~\tilde{\beta}_2(n)=\beta_2 -r_{1}(n) \, \beta_1 + \bigl(r_{2}(n)-r^2_{1}(n)\bigr)\, \beta_0 ,
\label{tb2}
\ee
with
\be
r_{2}(n)=\frac{C_{2}(n)}{d(n)}-\frac{d(n)-1}{2}\, r_1^{2}(n)\, .
\label{r2}
\ee

With the above choice of the scale, we have
\be
C_1(n)=
C_2(n) = 0,~~
\label{oBSI.NLO}
\ee
i.e. Eq.(\ref{oCSI.NLO}) is correct at the NNLO level:
\be
\overline{C}(n)= a_{2,n}^{d(n)}(Q^2)\, \Bigl[1+O(a_{2,n}^3)\Bigr]\, .
\label{oCSI.NNLO}
\ee

\section{Fitting procedure}

The most popular method (see, for example,~\cite{NNLOfits}) is to conduct QCD analysis over 
a wide range of data by using the Dokshitzer--Gribov--Lipatov--Altarelli--Parisi (DGLAP) integro--differential equations~\cite{DGLAP}.
This obviously is a brute force calculation that allows one to analyze the data directly.

At the same time, as can be seen from previous papers~\cite{PKK,KKPS1,KPS,Shaikhatdenov:2009xd,Kotikov:2015zda} there are
different approaches to the problem one of which was observed in~\cite{Barker} and developed in~\cite{Kri}.
This approach is based on the analysis of SF $F_2(x,Q^2)$ moments, which are actually solutions to DGLAP equations
in the Mellin moment space as defined in Eq.~(\ref{Moments}).
Then for each $Q^2$ SF $F_2(x,Q^2)$ is reconstructed using the Jacobi polynomial decomposition method \cite{Barker,Kri}:
\be
F_{2}(x,Q^2)=x^a(1-x)^b\sum_{n=0}^{N_{max}}\Theta_n ^{a,b}(x)\sum_{j=0}^{n}c_j^{(n)}(a,b)
M_{j+2} (Q^2)\,,
\label{2.1}
\ee
where $\Theta_n^{a,b}(x)$ denote Jacobi polynomials: $\Theta_n^{a,b}(x)=\sum_{j=0}^{n}c_j^{(n)}(a,b)\, x^j$,
while $a,b$ stand for the parameters to be fit.
As usual, the compliance condition is the requirement of error minimization while restoring the structure functions.

The program MINUIT \cite{MINUIT} is used to minimize the variable
\be
\chi^2_{SF} = \biggl|\frac{F_2^{exp} - F_2^{th}}{\Delta F_2^{exp}}\biggr|^2\,.
\label{chi2}
\ee

\section{Results}

We use free data normalizations for various experiments. The most stable BCDMS hydrogen data are used as a reference set
at the initial beam energy value $E_0=200$ GeV.
Unlike in our previous analyses~\cite{Shaikhatdenov:2009xd,Kotikov:2015zda}, the cut $Q^2\geq 2$ GeV$^2$
is used throughout, since for lower $Q^2$ values Eqs. (\ref{1.coA.DIS}) and (\ref{1.coB.DIS}) have no real solutions.

The starting point of the $Q^2$  evolution is the value of $Q^2_0$ = 90 GeV$^2$, which is close to the average  $Q^2$ value 
(on a logarithmic scale) of the data under study.
Based on previous investigations (see Ref. \cite{Kri}), the maximum number of moments used in the analyses is $N_ {max} = $8. The cut $0.25 \leq x \leq 0.8$ is also imposed on the data.

We work within the framework of the variable flavor number scheme (VFNS).
The threshold crossing point is taken at $Q^2_f=m^2_f$  (see \cite{Shaikhatdenov:2009xd}).
In order to emphasize the effect of changing the sign for twist four corrections, the results obtained in the fixed flavor number scheme (FFNS) with $n_f=4$ are shown as well.

Following our previous analysis carried out in \cite{Kotikov:2022vlo}, we expect a large $x$ resummation only slightly change
the strong coupling normalization, at the same time greatly modify the twist four values. Since the latter depend significantly on which data to be analyzed, here we will limit ourselves to dealing with exclusively hydrogen data.

In subsection 6.2 we examine the effect of resumming large $x$  logarithmic contributions using three different
resummation procedures discussed above. But first, we show the results of standard analysis and their dependence on the factorization $\mu_F$ and renormalization $\mu_R$ scales.

\vspace{0.5cm}
       {\bf Table 1.}  Twist four $\tilde h_4(x)$ parameter values obtained while fitting hydrogen data (total 314 points, $Q^2\geq 2$ GeV$^2$).
       Calculations are carried out within VFNS
       (FFNS). 
\footnotesize
       \begin{center}
\begin{tabular}{|l|c|c|c|}
\hline
&NLO &NNLO   \\
$x$
 &$\chi^2=246 (259)$ &$\chi^2=241 (254)$      \\
&$\alpha_s(M_Z^2)= 0.1195$ &$\alpha_s(M_Z^2)= 0.1177$   \\
&(0.1192) &(0.1170) \\
\hline \hline
0.275 & -0.25$\pm$0.02 (-0.26$\pm$0.03) & -0.19$\pm$0.02 (-0.20$\pm$0.02)   \\
0.35 & -0.24$\pm$0.02 (-0.25$\pm$0.02)  & -0.19$\pm$0.03 (-0.19$\pm$0.02)  \\
0.45 & -0.19$\pm$0.02 (-0.19$\pm$0.02) & -0.17$\pm$0.03 (-0.16$\pm$0.01)   \\
0.55 & -0.12$\pm$0.03 (-0.10$\pm$0.03)& -0.17$\pm$0.05 (-0.14$\pm$0.03)   \\
0.65 & 0.05$\pm$0.08 (0.12$\pm$0.08) &  -0.14$\pm$0.14 (-0.05$\pm$0.06)  \\
0.75 & 0.34$\pm$0.12 (0.48$\pm$0.12) & -0.11$\pm$0.19 (0.06$\pm$0.10)   \\
\hline
\end{tabular}
\end{center}
\vspace{0.5cm}
\normalsize

It is seen that the NS QCD analysis of SLAC, NMC and BCDMS experimental data for SF$F_2$ gives the following result
at the reference point:
\be
\as(M_Z^2) = 0.1177 +
\biggl\{\begin{array}{l}
\pm 0.0003 ~\mbox{(stat)}
\pm 0.0018 ~\mbox{(syst)}  \pm 0.0007 ~\mbox{(norm)} \\
\pm 0.0020 ~\mbox{(total exp.error)} \end{array}
\,.
\label{ConvAna}
\ee
Thus one can observe that these results look quite similar to those presented in \cite{Shaikhatdenov:2009xd,Kotikov:2015zda}.

\subsection{Scale dependence}

Let us study the dependence of the results on a different choice of factorization
$\mu_F=k_F Q^2$ and renormalization $\mu_R=k_R Q^2$ scales.
Following  \cite{Shaikhatdenov:2009xd,Kotikov:2015zda}, we select three values ($1/2,~1,~2$) for the coefficients
$k_F$ and $k_R$.

Results are demonstrated in Table~2.
The change in $\asMZ$ value for various $k_F$ and $k_R$ values is denoted by the difference:
\bea
\Delta \as(M_Z^2) ~=~ \as(M_Z^2) - \as(M_Z^2)|_{k_F=k_R=1}\,.
\eea

\vspace{4.5cm}
       {\bf Table 2.} NNLO (NLO) $\asMZ$ for a set of $k_F$ and $k_R$ coefficients,  (314 points, $Q^2\geq 2$ GeV$^2$). Calculations are carried out within VFNS.
\vspace{0.2cm}
\begin{center}
\begin{tabular}{|c|c||c|c|c|c|}
\hline
& & & &  \\
$k_R$ & $k_F$ & $\chi^2(F_2)$ & $\asMZ$ & $\Delta \as(M_Z^2)$ \\
& & & &  \\ \hline \hline
1   &  1   & 241 (246) & 0.1177 (0.1195)   &  0 \\
1/2 &  1   & 241 (246) & 0.1166 (0.1171)   & -0.0011 (-0.0024) \\
1   &  1/2 & 239 (243) & 0.1170 (0.1170)   & -0.0007 (-0.0025) \\
1   &  2   & 244 (249) & 0.1193 (0.1227)   & +0.0016 (+0.0032)  \\
2   &  1   & 243 (247) & 0.1191 (0.1225)   & +0.0014 (+0.0030) \\
\hline
\end{tabular}
\end{center}

As can be seen from this table, the theoretical uncertainties for the maximum and minimum values of the coupling constant
corresponding to $k_i=2$ and $k_i=1/2$ ($i=F,R$), respectively, are equal to $+0.0021$ ($+0.0044$) and $-0.0013$ ($-0.0035$) for the case of NNLO (NLO)
\footnote{Note that here we take the theoretical errors for factorization and renormalization scales in quadrature.
  In our previous analyses \cite{Shaikhatdenov:2009xd,KK2001}, we considered cases with $k_F=k_R=1/2$ and $k_F=k_R=2$,
  which corresponded to summing up corresponding errors linearly rather than in quadrature.}.
It should be noted that we take into account the uncertainty of the renormalization scale in the expressions for coefficient
functions and corresponding coupling constants in a way similar to what was done in~\cite{NeVo}.

Thus, the present analysis gives the following theoretical error for the result presented in (\ref{ConvAna}):
\be
\as(M_Z^2) = 0.1177
\pm 0.0020 ~\mbox{(total exp.error)}
+ \biggl\{
\begin{array}{l}+ 0.0021 \\ -0.0013 \end{array} ~\mbox{(theor)}\,.
\label{ConvAna+FR}
\ee

\subsection{Resummation}

Now we repeat NS QCD analysis performed in this Section above (whose results are shown in Table~1), this time by using the schemes that
contain effective resummation of large $x$ logarithms, which, in turn, contribute to the Wilson coefficient functions.
These schemes are the DIS-scheme \cite{DIS}, $W^2$-evolution and the Grunberg effective charge method \cite{Grunberg},
which are presented in detail in Section~4 above.

\vspace{0.5cm}
       {\bf Table 3.} Same as in Table~1 but
       carried out within VFNS only      
\footnotesize
       \begin{center}
\begin{tabular}{|l|c|c|c|c|}
\hline
&NLO DIS scheme ($W^2$-evolution)[SIPT]
&NNLO DIS scheme ($W^2$-evolution)[SIPT]\\
$x$ &$\chi^2=238 (251) [245]$  &$\chi^2=242 (249) [249]$     \\
&$\alpha_s(M_Z^2)= 0.1177 (0.1179)[0.1178]$ & $\alpha_s(M_Z^2)= 0.1160 (0.1163)[0.1178]$    \\
\hline \hline
0.275 & -0.18$\pm$0.01 (-0.17$\pm$0.02) [-0.22$\pm$0.03] & -0.14$\pm$0.01 (-0.13$\pm$0.03) [-0.17$\pm$0.02]  \\
0.35 & -0.11$\pm$0.01 (-0.13$\pm$0.01) [-0.15$\pm$0.02]&   -0.13$\pm$0.02 (-0.08$\pm$0.02) [-0.14$\pm$0.02]  \\
0.45 & -0.04$\pm$0.04 (-0.09$\pm$0.01) [-0.07$\pm$0.03]  & -0.11$\pm$0.09 (0.02$\pm$0.02)  [-0.10$\pm$0.02]   \\
0.55 & -0.10$\pm$0.01 (-0.09$\pm$0.04)  [-0.12$\pm$0.03] &  -0.12$\pm$0.03 (0.08$\pm$0.03) [-0.09$\pm$0.02] \\
0.65 & -0.17$\pm$0.04 (-0.09$\pm$0.05) [-0.14$\pm$0.05] &  -0.22$\pm$0.05 (0.08$\pm$0.05) [-0.12$\pm$0.04]  \\
0.75 & -0.57$\pm$0.08 (-0.46$\pm$0.18) [-0.51$\pm$0.09] &  -0.59$\pm$0.08 (-0.18$\pm$0.10)  [-0.42$\pm$0.09]  \\
\hline
\end{tabular}
\end{center}
\vspace{0.5cm}
\normalsize

\vspace{3.5cm}
       {\bf Table 4.}  Same as in Table 1 but
       carried out within FFNS only 
\footnotesize
       \begin{center}
\begin{tabular}{|l|c|c|c|c|}
\hline
&NLO DIS scheme ($W^2$-evolution)[SIPT]
&NNLO DIS scheme ($W^2$-evolution)[SIPT]\\
$x$ &$\chi^2=244 (243) [245]$  &$\chi^2=244 (245) [245]$     \\
&$\alpha_s(M_Z^2)= 0.1180 (0.1179)[0.1179]$ & $\alpha_s(M_Z^2)= 0.1160 (0.1163)[0.1171]$    \\
\hline \hline
0.275 & -0.21$\pm$0.03 (-0.19$\pm$0.03) [-0.22$\pm$0.03] & -0.22$\pm$0.03 (-0.13$\pm$0.03) [-0.22$\pm$0.01]  \\
0.35 & -0.13$\pm$0.02 (-0.11$\pm$0.02) [-0.15$\pm$0.02]&   -0.14$\pm$0.02 (-0.08$\pm$0.02) [-0.12$\pm$0.01]  \\
0.45 & -0.05$\pm$0.03 (-0.04$\pm$0.03) [-0.07$\pm$0.03]  & -0.03$\pm$0.03 (0.02$\pm$0.02)  [0.02$\pm$0.02]   \\
0.55 & -0.10$\pm$0.03 (-0.10$\pm$0.03)  [-0.12$\pm$0.03] &  -0.03$\pm$0.03 (0.08$\pm$0.03) [0.01$\pm$0.04] \\
0.65 & -0.11$\pm$0.05 (-0.15$\pm$0.05) [-0.14$\pm$0.05] &  0.00$\pm$0.05 (0.08$\pm$0.05) [0.05$\pm$0.09]  \\
0.75 & -0.45$\pm$0.09 (-0.52$\pm$0.09) [-0.51$\pm$0.09] &  -0.32$\pm$0.09 (-0.18$\pm$0.10)  [-0.26$\pm$0.15]  \\
\hline
\end{tabular}
\end{center}
\vspace{0.5cm}
\normalsize
  It can be seen from Tables 1--4 that upon switching to schemes that effectively take into account resummations in the region
  of large $x$ values, the normalized value of the strong coupling constant remains almost intact whereas the form of
  twist four corrections varies noticeably.

  Indeed, used resummation schemes change slightly the twist four terms in the area of relatively small $x$ values while
  these corrections in the region of large $x$ values change their sign (see also Fig.~1). These changes in the values of
  twist four corrections are
  nearly independent of both the chosen resummation scheme and the order of perturbation theory.
Moreover, it seems that they rise as $1/(1-x)$ at large $x$ but this observation needs additional investigations.

\begin{figure}[!htb]
\centering
\includegraphics[width=0.58\textwidth]{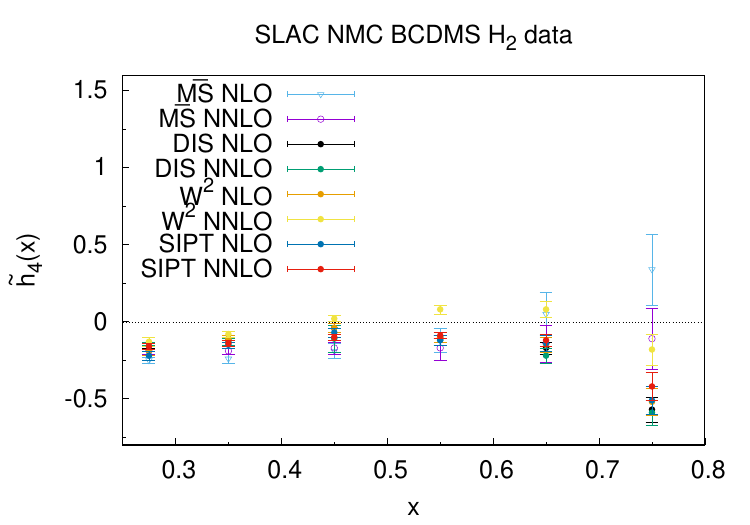}
    \caption{\label{fig:APTHT}
The values of the parameters $\tilde{h}_4(x)$ obtained in the analysis of experimental data within VFNS.
    }
\end{figure}

Such a behavior is in sharp contrast with the analyses
\cite{Shaikhatdenov:2009xd,Kotikov:2015zda,Blumlein:2008kz,Alekhin:2000ch} performed in
${\mathbf{\overline{MS}}}$ scheme, where twist four corrections are mostly positive at large $x$ and rise as $1/(1-x)$
(see also Table~1 and Fig.~1).

Negative values of twist four corrections for large $x$, obtained in schemes with resummation of large $x$ logarithms, can lead
to the following phenomenon: at least part of the (negative) power terms can be absorbed by the difference between the
usual strong coupling and the analytic one \cite{SoShi}, if we use the analytic coupling constant
\footnote{The analytic coupling constant was recently obtained in high orders of perturbation theory
  in Ref. \cite{Kotikov:2022sos}.}
in our analyses.
This would happen exactly as it was the case at low $x$ values (see  Refs.~\cite{CIKK09,Kotikov:2012sm}) within the framework of
the so-called double asymptotic scaling approach \cite{Q2evo}.
Of course, such a phenomenon was absent in the case of the ${\mathbf{\overline{MS}}}$ scheme, where the use of analytic QCD coupling~\cite{Kotikov:2010bm}
simply increases the magnitude of the twist four corrections.

Note that in previous works (see \cite{Vovk,PKK}), where resummation at large $x$ values was performed within the framework
of the Grunberg approach \cite{Grunberg}, only decrease in the twist four contributions was observed, since the
corresponding contributions were not studied in detail.

\section{Summary}

We analyze the experimental data collected by BCDMS, SLAC and NMC collaborations for DIS SF $F_2(x,Q^2)$ by resumming large
logarithms at large $x$ values into the corresponding Wilson coefficient function. For this matter we apply three schemes:
the DIS scheme \cite{DIS}, the $W^2$ evolution and the Grunberg effective charge method \cite{Grunberg}, which are
presented and studied in detail in Sections 3 and 4 above.

It is shown that the use of schemes with effective resummation of large logarithms at large $x$ values does not visibly change the strong coupling constant $\alpha_s(M_Z^2)$ values; however, those of the twist four corrections become
large and negative, which contradicts the results obtained within ${\mathbf{\overline{MS}}}$ scheme.

It seems that the negative values of twist four corrections obtained for large $x$ can be absorbed by the
difference between the
usual and analytic strong coupling constant~\cite{SoShi} provided we use the analytic one in the calculations.
Let us hope that such kind of analysis will be performed sometime in the future.

\section{Acknowledgments}

One of us (A.V.K) was supported in part by the Foundation for the
Advancement of Theoretical Physics and Mathematics BASIS.  He thanks the Sun Yat-sen University School of Physics
and Astronomy for the invitation.

\end{document}